\documentclass[universe,review,accept,moreauthors,pdftex]{Definitions/mdpi}
\graphicspath{{Figures/}}
\usepackage{diagbox}
\usepackage{amsmath, amssymb}
\firstpage{1}
\pubvolume{7}
\issuenum{12}
\articlenumber{497}
\pubyear{2021}
\copyrightyear{2021}
\externaleditor{Academic Editors: {Philippe Jetzer and Domenico Giulini} 
} 
\datereceived{21 October 2021}
\dateaccepted{8 December 2021}
\datepublished{16 December 2021}
\hreflink{https://\\doi.org/10.3390/universe7120497} 
\usepackage{color}

\Title{Testing General Relativity with Gravitational Waves: An~Overview}
\TitleCitation{Testing General Relativity with Gravitational Waves: An Overview}

\Author{{N.~V.~Krishnendu} $^{1,2,}${*}\href{https://orcid.org/0000-0002-3483-7517}{\orcidicon} and Frank Ohme $^{1,2}$\href{https://orcid.org/0000-0003-0493-5607}{\orcidicon}}

\AuthorNames{N V Krishnendu and Frank Ohme}

\AuthorCitation{Krishnendu, N.V.; Ohme, F.}

\address{%
$^{1}$ \quad Max Planck Institute for Gravitational Physics (Albert Einstein
Institute), Callinstr. 38, \mbox{D-30167 Hannover, Germany}; frank.ohme@aei.mpg.de
\\
$^{2}$ \quad  {Institut f\"ur Gravitationsphysik, Leibniz Universit\"at Hannover}, D-30167 Hannover, Germany }

\corres{Correspondence: krnava@aei.mpg.de}

\abstract{The detections of gravitational-wave (GW) signals from compact binary coalescence by ground-based detectors have opened up the era of GW astronomy. These observations provide opportunities to test Einstein's general theory of relativity at the strong-field regime. Here we give a brief overview of the various GW-based tests of General Relativity (GR) performed by the LIGO-Virgo collaboration on the detected GW events to date. After providing details for the tests performed in four categories, we discuss the prospects for each test in the context of future GW detectors. The four categories of tests include the consistency tests, parametrized tests for GW generation and propagation, tests for the merger remnant properties, and GW polarization tests.  }


\begin{document}

\def \Bilby{\texttt{Bilby}\xspace}
\def \MAP{\texttt{MAP}\xspace}
\def \lalinference{\texttt{LALInference}\xspace}
\def \PhenomPvthreeHM{\texttt{IMRPhenomPv3HM}\xspace}
\def \PhenomHM{\texttt{IMRPhenomHM}\xspace}
\def \msun{$\rm{M}_{\odot}$\xspace}
\def \PhenomPvthree{\texttt{IMRPhenomPv3}\xspace}

\newcommand{\fo}[1]{\textcolor{red}{[Frank: #1]}}



\section{{Introduction}}

The binary evolution in General Relativity (GR) is described differently than in Newtonian gravity (NG). In GR, the binary orbit shrinks due to the emission of energy: angular and linear momenta through gravitational waves (GWs)~\cite{Kepler:1609, Newton:1687, Einstein:1918, Einstein:1916}. Whereas in NG, there is no concept of radiation reaction and the orbital period is constant over time. Even though Albert Einstein predicted the existence of GWs more than a century ago, their detection remained a puzzle due to their weak interaction with matter. The indirect evidence of GWs came from the decades-long observations of orbital decay of a binary pulsar by Russell Alan Hulse and Joseph Taylor~\cite{Hulse1975Discovery, Taylor:1994, Blanchet:2003uf, Burgay03}.  They found that the observed orbital decay of the binary system, known as PSR B1913+16(PSR J1915+1606, or PSR 1913+16), due to the emission of GWs, is consistent with the predictions of GR. That is, the rate of decay of the orbital period ($P_{orb}$) from observations $\dot{P}_{orb}\sim10^{-14}$--$10^{-12}$ agreed to the GR predicted rate obtained from analytical calculations based on GR, leading the team to win the Physics Nobel Prize in 1993.

The direct detection of GWs had to wait until the LIGO detectors at Hanford, Washington, and Livingston, Louisiana, made their first detection of a binary merger on 14~September 2015~\cite{TOGGW150914, PEGW150914, AstroGW150914, gw150914, Advanced_LIGO_Reference_Design,AdvancedLIGO,TheVirgostatus,TheVirgo:2014hva,LIGOScientific:2014pky}. This discovery opened the era of GW astronomy, noting the first highly relativistic strong-field observation of GWs. Within the subsequent years of observation runs, the LIGO-Virgo collaborations announced the detection of more than fifty binary merger events~\cite{GWTC-2LIGOScientific:2020ibl,gw150914,GW151226,GW170104,GW170608,GW170814,GW170817,O1O2catalogLSC2018,GW190425,GW190412,GW190814, GW190521,IAScatalog}.

\textls[-15]{Among the many significant contributions to fundamental physics and astrophysics, GW observations test GR at the relativistic, strong-field regime. A set of testing GR analyses conducted by the LIGO-Virgo Scientific Collaboration (LVC) on the {\tt GW150914} event established that {\tt GW150914} is consistent with a binary black hole (BBH) signal described in GR~\cite{TOGGW150914}.}

This set of tests include consistency tests, parameterized tests, tests to confirm the non-dispersive nature of the radiation, and tests on the remnant properties. Consistency tests check for the agreement of observed data with the signal predicted from GR. Parameterized tests introduce model-agnostic parametric GR deviations in the waveform and constrain those from the data to put statistical bounds on these parameters. The list of the GW events was extended further with more binary merger detections by LIGO and Virgo detectors' first, second, and third observation runs. All the tests applied on {\tt GW150914} were also performed on these events with appropriate modifications to the above-mentioned tests, including additional tests. Here we will go through them in detail.

The GW-based tests of GR on the BBH coalescence events detected by LIGO and Virgo until 1 October 2019 are available in Reference~\cite{GWTC2-TGR}. A generic binary system evolves from its early inspiral weak-field regime to a highly relativistic merger and then the final ringdown stage. In the case of BBHs in GR, the object formed after they merge (i.e., the merger remnant) is another black hole (BH). On the other hand, for non-BH binaries~\cite{Cardoso:2017cfl,CardosoLivingReview},{\endnote{{Non-BH binaries include other compact objects like neutron stars and more exotic objects like boson stars~\cite{Jetzer:1991jr,Liddle:1993ha,Schunck:2003kk,Liebling:2012fv,Brito:2015pxa,Friedberg:1986tp,Friedberg:1986tq,Ryan97b,Visinelli:2014twa}, gravastars~\cite{Chirenti:2016hzd,Chirenti:2007mk,Lobo:2005uf,Carter:2005pi,Visser:2003ge,Mazur:2001fv} etc.}}} the merger remnant is not necessarily a BH but could be another compact star depending upon the properties of the binary.

\begin{figure}[H]
\includegraphics[scale=0.4]{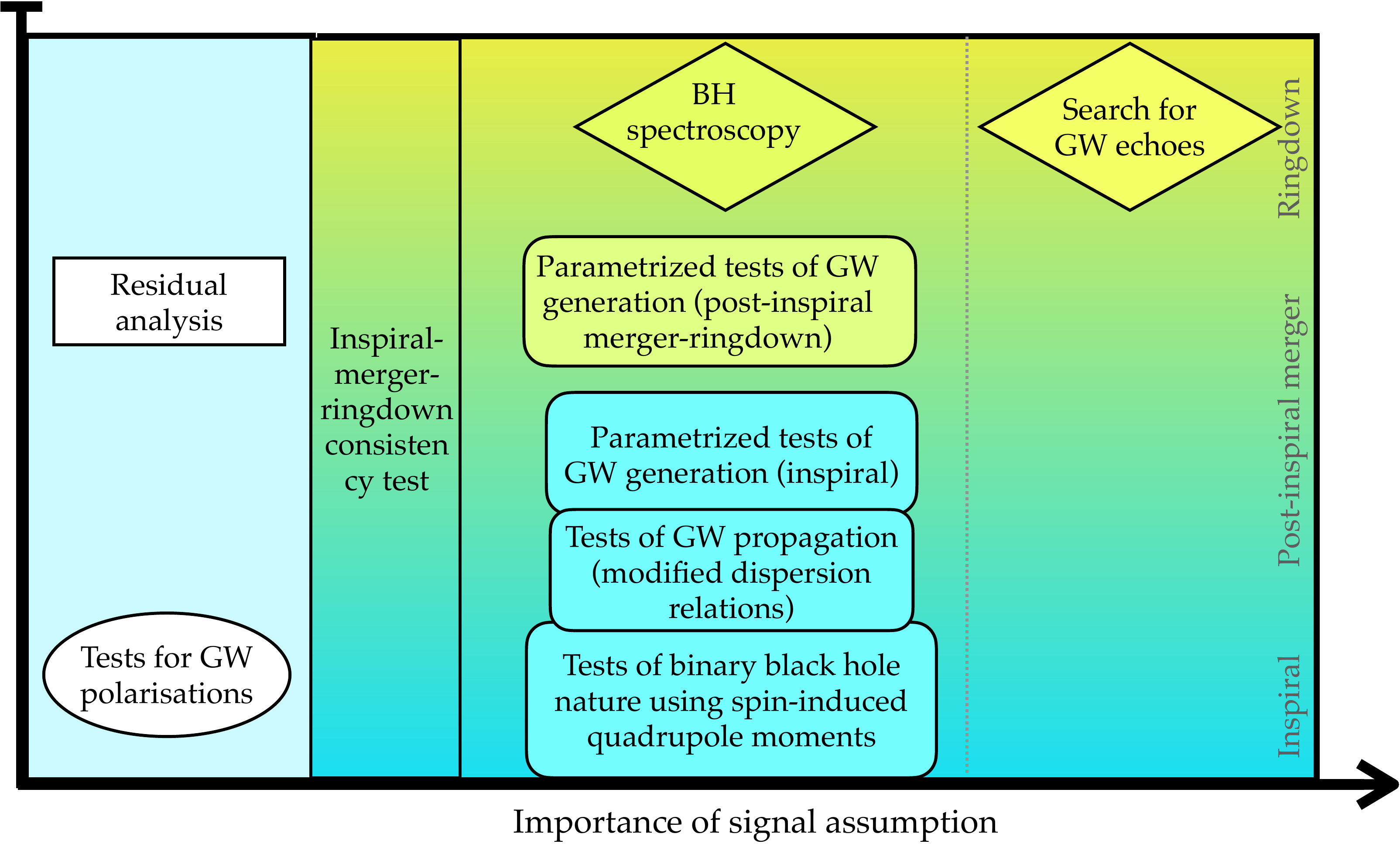}
\caption{Outline of various tests of GR we discuss in this article. The x-axis denotes the increasing order of GR model assumptions that go into each analysis. The light-blue region on the left side contains the set of tests that require the least assumptions about the signal model. The gradient on the right part of the plot classifies the tests into inspiral, merger-ringdown regimes of the signal from bottom to top. Different shapes indicate different classes of tests: rectangle, rounded rectangle, ellipse, and diamond shapes correspond to consistency tests, parametrized tests, polarization tests, and tests for the merger remnant, respectively. }
\label{fig:flowchart}
\end{figure}

In a model agnostic way, there were four broad classes of tests conducted in Reference~\cite{GWTC2-TGR}. These tests aim to look at different regimes of binary evolution or to the full inspiral-merger-ringdown signals. The first set consists of the residual analysis and the inspiral-merger-ringdown consistency test. Both of these tests check the consistency of GR predictions with the observed data (as in the case of Reference~\cite{TOGGW150914}). The second category of tests is the parametrized tests for GW generation and propagation. Here one sets statistical bounds on the parametrized deviations from GR, assuming GR is the correct theory of gravity, employing GR waveform models with parametric deviations present.  On the third category of tests, one looks for any violation of GR by analyzing the merger remnant properties. The GW polarization tests look for extra polarization modes present in the data and comprise the fourth set of tests. This analysis provides statistical evidence for alternative theories of gravity that predict vector and scalar polarization modes along with the tensor modes. An overview of these tests is provided in Figure~\ref{fig:flowchart}.

The probability of there being astrophysical origin of a  candidate event plays an important role in determining whether that event is considered for the testing GR analyses or not. Usually, a higher threshold is assumed so that the events analyzed have higher chances of being of astrophysical origin. For instance, in Reference~\cite{GWTC2-TGR}, events satisfying a false alarm rate (FAR) less than $10^{-3}$ per year are chosen to analyze. Once the set of events is chosen based on the detection significance, additional criteria are applied depending upon the strategies followed by each test. Bayesian formalism-based techniques are employed to get meaningful bounds from each test. The pipelines widely used for this purpose are, \mbox{{\tt LALInference} ~\cite{Veitch:2014wba}} available in the LIGO Scientific Collaboration's algorithm library suite ({\tt LALSuite})~\cite{LALSuite}, {\tt Bayeswave}~\cite{Cornish:2014kda,Littenberg:2014oda}, parallel bilby ({\tt pBilby})~\cite{Romero-Shaw:2020owr,Ashton:2018jfp,Smith:2019ucc}, {\tt{PYRING}}~\cite{PhysRevD.99.123029,Isi:2019aib} and {\tt Bantam}~\cite{Pang:2020pfz}. Reference~\cite{GWTC2-TGR} demonstrated the possibility of performing tests of GR on binary black hole (BBH) events, employing mainly two different waveform models, {\tt{IMRPhenomPv2} }(phenomenological waveform model for a precessing BBH \mbox{system)~\cite{Husa:2015iqa, Khan:2015jqa, Hannam:2013oca} and {\tt SEOBNRv4\_{ROM}}} (reduced-order effective one body (EOB, waveform model for a non-precessing binary system)~\cite{Bohe:2016gbl}.

There is a significant increase in the detection rate as the detectors improve their sensitivities through first, second, and third observing runs of LIGO-Virgo detectors. Interestingly, it is possible to infer information from multiple events by combining the data from each event. The combined bounds help to improve our understanding of binary population properties in general. As we combine results, the statistical uncertainty that arises due to instrumental noise lessens. Notice that this instrumental noise does not include the uncertainty contributions from the systematic errors of gravitational waveform modeling~\cite{LiEtal2011,Arun:2012hf,Sampson:2013lpa,Kramer:2006nb,Yunes:2010qb,Mirshekari:2011yq,LSC_waveform_model_systematicsGW150914}, calibration of the detectors, and power spectral density (PSD) estimation uncertainties~\cite{2017PhRvD..96j2001C,Viets:2017yvy,Sun:2020wke,LIGOScientific:2018kdd,Vajente:2019ycy}. Sometimes, systematic errors can dominate the statistical errors and lead to false identification of GR violations, which we do not discuss here. Previous studies in References~\cite{TOGGW150914, TGRGWTC-1, LSC_waveform_model_systematicsGW150914} discuss two different statistical approaches to estimate combined information on GR test parameters from multiple events. The first one (also called {restricted} or {simple} combining) assumes equal GR deviations across all the events independent of the physical parameters characterizing the binary, and this technique is well described and demonstrated for GWTC-1 events ~\cite{TGRGWTC-1}. This assumption is generally incorrect as there are cases when the waveform model can arbitrarily deviate from GR depending upon the binary source properties. The second method, the hierarchical combining strategy, tries to overcome the issue of universality assumption by relaxing it. In this case, instead of assuming uniform GR deviation for all events, a Gaussian distribution models the non-GR parameter. The statistical properties (mean ($\mu$) and standard deviation ($\sigma$)) of this distribution are obtained from the data itself, and the estimates are different for different models of gravity. We call the parameters $\mu$ and $\sigma$ hyperparameters. If GR is the correct theory, the Gaussian distribution should center around zero. The astrophysical population properties of sources play a crucial role in estimating these statistical quantities~\cite{GWTC2-TGR}.

The organization of this article is as follows. Section~\ref{tests_gr} is dedicated to various tests of GR performed already on the GW events observed by the LIGO-Virgo detectors, including tests of consistency with GR (Section~\ref{consistency_tests}), parametrized tests (Section~\ref{param_tests}), tests based on the merger remnant properties (Section~\ref{rem_properties}), and tests for GW polarizations (Section~\ref{pol}). We conclude with a summary section, Section~\ref{con}.

\section{Model-Agnostic Tests of General Relativity from Gravitational-Wave Observations}
\label{tests_gr}

\subsection{Tests of Consistency with General Relativity}
\label{consistency_tests}

Consistency tests do not need to assume any particular alternative theory to GR, nor do they test specific deviations. They address the simple question: can the observed data be fully explained by assuming GR? Or put differently, is there any statistically significant ``trace'' in the data that is unlikely to be explained as either part of an astrophysical signal (assuming GR) or the terrestrial instruments' noise?
So far two different kinds of consistency tests have been performed on the detected GW events~\cite{GWTC2-TGR}: the residual analysis and the Inspiral-merger-ringdown (IMR) consistency test.


\subsubsection{Residual Test}

The residual test checks for signatures left in the data after subtracting the best-fit GR template. If GR is the correct theory and we have subtracted the astrophysical signal completely, the residuals in each detector should be consistent with instrumental noises.

The best fit model of the astrophysical signal is obtained by a detailed parameter estimation analysis using a stochastic sampling of the signal's parameter space. Typically, the best-fit parameters are taken to be those that maximize the likelihood of observing the recorded data assuming this signal is present in the data. This set of parameters is not necessarily the one that describes the most probable source configuration a posteriori, as the likelihood alone does not take prior assumptions into account. Nevertheless, the maximum-likelihood parameters are those that minimize the difference between the data $d_i$ and template $h_i$ by definition of the likelihood $\Lambda$,
\begin{equation}
\log \Lambda = - \frac 1 2 \sum_{i} \| d_i - h_i \|^2. \label{eq:loglikelihodd}
\end{equation}

{Here}, the index $i$ enumerates the different detectors; $h_i$ is the signal projected onto each detector, respectively, and the norm $\| \cdot \|^2 = \langle \cdot, \cdot \rangle$ is induced by the following inner~product
\begin{equation}
\left \langle a, b \right \rangle = 4 \Re \int \frac{\tilde a(f) \, \tilde b(f)^\ast}{S_n(f)} \; df. \label{eq:inner_prod}
\end{equation}

{The} detector noise spectral density $S_n(f)$ acts as a weight in an integral over the Fourier transformed functions $\tilde a$ and $\tilde b$; $^\ast$ denotes complex conjugation.

By this construction, the residuals $d_i - h_i$ are small in the sense of Equation~(\ref{eq:loglikelihodd}), but they could still contain a coherent signal that cannot be captured by the GR model. To look for such a potential signal, the method employed in References~\cite{GWTC2-TGR,TOGGW150914, LSC_waveform_model_systematicsGW150914} is {\tt Bayeswave}~\cite{Cornish:2014kda}: a transient search algorithm looking for coherent excess power in the (residual) detector data. This part of the analysis is model-independent. {\tt Bayeswave} uses Morlet--Gabor wavelets to look for coherent, elliptically polarized signatures that rely on no further model assumption. In addition to generic signals, it employs models for stationary and non-stationary noise simultaneously.

To quantitatively explain the results, we require various definitions. First, the optimal network signal-to-noise-ratio (SNR) of a signal $h$ is derived from its norm,
\begin{equation}
\textrm{SNR}(h) = \sqrt{\sum_i \langle h_i, h_i \rangle } = \sqrt{\sum_i \| h_i \| ^2},
\end{equation}
using the inner product Equation~(\ref{eq:inner_prod}). If we take $h$ as the best-fit GR template, we obtain $\textrm{SNR}_{\rm GR}$. The residual modelled by {\tt Bayeswave} is not a single signal. The uncertainty in what constitutes a coherent residual signal and what is instrument noise leads {\tt Bayeswave} to provide a discretized probability distribution in the parameter space of wavelets. However, each point in this distribution corresponds to a residual signal that has a well-defined SNR. Consequently, we can map the probability distribution of residual signals to a probability of their optimal network SNRs.

As is standard, we characterize the probability distribution by \emph{credible intervals} that enclose a certain amount of probability (we use this quantity more frequently in this article). Specifically, we report the SNR of the residual at which the cumulative probability distribution is 90\%. Put differently, we infer a 90\% probability that the residual signal after subtracting the best-fit GR template has an optimal network SNR $\leq \textrm{SNR}_{90}$. 

\begin{figure}[H]
\widefigure
\includegraphics[width=0.96\textwidth]{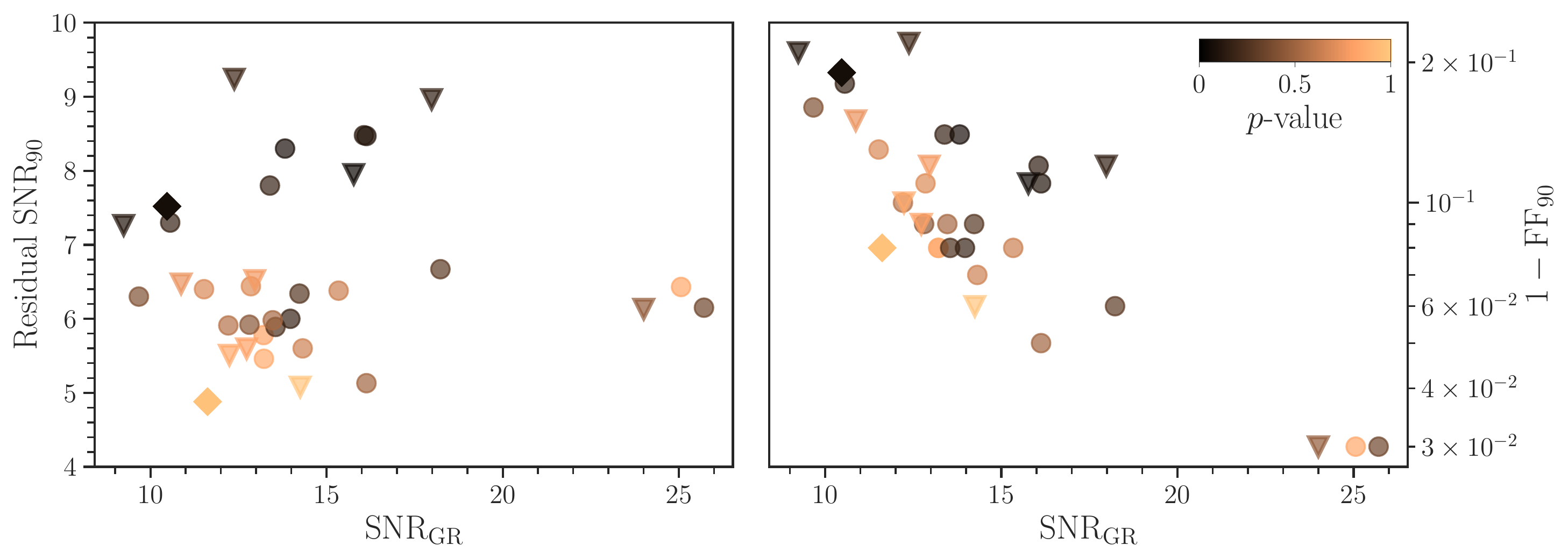}
\caption{{The} residual SNR (\textbf{left}) and the fitting factor (\textbf{right}) as functions of the SNR of the best-fit GR model for BBH observations reported in \cite{GWTC2-TGR}. The colorbar indicates the  $p$-value associated with each event, with diamond markers noting the maximum and minimum values (see text). O3a events are distinguished from the O1/O2 events by circle and triangle-shaped markers, respectively.}
\label{fig:res1}
\end{figure}

The left panel of Figure~\ref{fig:res1} shows $\textrm{SNR}_{90}$ as a function of the best-fit $\textrm{SNR}_{\rm GR}$ for the observed binary mergers from O1, O2, and O3a. The SNR of the GR signals ranges between 9.24 ({\tt GW151012}) and 25.71 ({\tt GW190521}). The 90\% upper credible bound of the residual SNR ranges between 4.88 ({\tt GW190727\_060333}) and 9.24 ({\tt GW170818}). No clear correlation is visible between the two quantities. If our GR models would consistently be unable to capture an ubiquitous deviation from GR, we might expect that stronger signals correlate with stronger residuals. The current data shows no indications of such correlations.

\end{paracol}
\nointerlineskip

\begin{paracol}{2}
\switchcolumn
To systematically assess if the residual SNRs are consistent with the detector noise, we can define a $p$-value under the hypothesis that the residual is consistent with detector noise (this is the null hypothesis). This $p$-value is estimated by running identical {\tt Bayeswave} analyses on a large number of noise-only data segments around, but not including, each event. The $p$-value then provides the probability of pure noise producing an $\textrm{SNR}^{n}_{90}$ greater than or equal to the residual SNR found after subtracting the best-fit GR template, $\textrm{SNR}_{90}$. That implies, $p$-value$:=P(\textrm{SNR}_{90}^{n}\geq \textrm{SNR}_{90})$. A large $p$-value indicates that there is a high chance that the residual power originates from the instrumental noise. Small $p$-values, on the other hand, indicate that it is less likely for noise alone to yield such high values of residual SNR.

The $p$-values for all events considered here are included in Figure~\ref{fig:res1} as a color scale. They also span a large range between $0.07$ ({\tt GW90421\_213856}) and $0.97$ ({\tt GW190727\_060333}). This is to be expected in repeated uncorrelated experiments. In fact, assuming the residuals are pure noise, the $p$-values of the population should be uniformly distributed between $[0, 1]$. As further discussed and quantified in \cite{GWTC2-TGR}, the $p$-values found for O1, O2, and O3a events are consistent with this expectation.

As a final interpretation of the residuals, one can ask: how well does the GR model fit the signal in the data? Obviously, if the model would be perfect, and we could unambiguously identify the coherent signal in the data, then the agreement between the model and data should be perfect, too, assuming GR is the correct theory. In reality, our GR models are very accurate, but not perfect, and we only have a probabilistic measure of the signal in the data. Therefore, we can only expect to obtain a lower bound on the \emph{fitting factor} $\mathrm{FF}$ between the model $h$ and the signal $s$ by calculating
\begin{align}
\mathrm{FF} &= \frac{\langle h, s \rangle}{\| h \| \| s \|} = \frac{\langle h, h + s_r \rangle}{\| h \| \| h + s_r \|} = \frac{\| h \|}{\sqrt{\| h \|^2 + \| s_r \|^2}} \\
\Rightarrow ~~ \mathrm{FF}_{90} & = \frac{\mathrm{SNR}_{\rm GR}}{\sqrt{\mathrm{SNR}_{\rm GR}^2 + \mathrm{SNR}_{90}^2}}.
\end{align}

{Here} we used that the coherent signal $s$ is the sum of the GR model $h$ and any residual $s_r$ that is perpendicular to $h$. The latter assumption is justified because $h$ was chosen by maximizing the agreement between the data and the model. $\mathrm{FF}_{90} = 1$ would indicate perfect agreement.

We plot $1 - \mathrm{FF}_{90}$ on the right panel of Figure~\ref{fig:res1} for the observations we considered so far. They show more clearly a correlation with $\mathrm{SNR}_{\rm GR}$. Strong signals with large values of $\mathrm{SNR}_{\rm GR}$ tend to yield higher fitting factors. This is not because the models describe the actual signal better for louder events. It is because our confidence in how well the model agrees with the actual signal increases with increasing $\mathrm{SNR}_{\rm GR}$.

Another notable fact about the fitting factors is that they are larger than related quantities one often finds in the GW literature. For example, discrete template banks for GW searches are often constructed such that fitting factors of at least $0.97$ are guaranteed between any signal and the closest template in the bank \cite{LIGOScientific:2016vbw}. Waveform models for BBHs are commonly tuned to $\mathrm{FF} \lesssim 10^{-3}$ between the most accurate predictions and the full model \cite{Ohme:2011zm, Ohme:2011rm}. Parameter estimation poses strict demands on waveform accuracy of the order of $\mathrm{SNR}_{\rm GR}^{-2}$ \cite{Lindblom:2008cm, Baird:2012cu} (i.e., waveform differences of $< 10^{-2}$). Is that level of accuracy necessary, given that for most observations, we cannot put stronger constraints on the fitting factor than $\mathrm{FF}_{90} \sim \mathcal O (10^{-1})$? The answer is, of course, yes! While we cannot be sure about the true signal for individual events, it is worth emphasizing that the {\tt Bayeswave} analysis has great freedom and a large parameter space to identify virtually {any} residual signature as a potential coherent signal. Much more restricted measurements that only look for specific, lower-dimensional deviations are sensitive to significantly smaller signal differences, because only those signal differences that are consistent with both the assumed variation and the data are considered. Therefore, the residuals test is a very generic baseline test for anything that cannot be modeled with GR. It is, however, much less sensitive to specific deviations that can be tested more accurately in a dedicated test (see the rest of this paper). A more detailed discussion of the relation between various tests of GR can be found in \cite{Johnson-McDaniel:2021yge}. 


\subsubsection{Inspiral-Merger-Ringdown Consistency Test}

In GR, the time evolution of BBH mergers is uniquely determined. Hence the final mass and spin of the remnant BH are uniquely determined from the initial mass and spin parameters of Kerr BHs. The inspiral-merger-ringdown consistency test (IMRCT) tests the consistency of inspiral and merger-ringdown parts of the signal by comparing two independent estimates of binary parameters. More specifically, the binary's final mass and spin parameters are measured separately from both low- and high-frequency parts of the signals and then compared to the two measurements to check their agreement.

Given the final mass $(M_{f})$ and spin $(\chi_{f})$, one can estimate the spin-dependent innermost stable circular frequency\endnote{Inner-most stable circular orbit of a Kerr BH is the smallest stable circular orbit in which a test particle can stably orbit around the~BH.} for a Kerr BH ($f_{\rm{cut}}$)~\cite{Jimenez-Forteza:2016oae,Hofmann_2016,Healy:2016lce}. The full BBH signal can be divided into two parts using this frequency, the low-frequency part and the high-frequency part{\endnote {Current analysis is taking into account for the dominant mode and neglecting any higher-mode contributions to the frequency evaluation.}}. By restricting the noise-weighted integral in the likelihood calculation to frequencies below $(f<f_{\rm{cut}})$ and above this frequency cutoff $(f>f_{\rm{cut}})$, the binary parameters are estimated using stochastic sampling algorithms based on Bayesian inference. The merger remnant properties are calculated by averaging NR-calibrated final state fits given the posterior median values~\cite{Jimenez-Forteza:2016oae, Hofmann_2016, Healy:2016lce} from the two independent mass-spin estimates above. This calculation assumes an aligned-spin binary system.  If the data is consistent with GR, both estimates should agree~\cite{Ghosh:2018,  PhysRevD.90.104004, gw150914, GW170104}.

The frequency, $f_{\rm{cut}}$, roughly divides the signal into inspiral and merger-ringdown (post-inspiral) regimes. To calculate $f_{\rm{cut}}$, the binary parameters inferred from the full signal are used. As the test relies on independent parameter inference from the low and high frequency parts of the full signal, one requires enough SNR in both these regimes of the signal. For the selected events, a detailed parameter estimation analysis is performed in Reference~\cite{GWTC2-TGR} focusing on the mass-spin parameters. If $M_{f}^{insp}$ and $M_{f}^{post-insp}$ denote the final mass estimates obtained from low and high frequency parts of the signal, we can define a dimension-less quantity that measures the fractional deviation from the final-mass estimate~as,
\begin{equation}
\delta M_{f} = \frac{\Delta M_{f}}{\bar{M_{f}}}=2\frac{M_{f}^{insp}-M_{f}^{post-insp}}{M_{f}^{insp}+M_{f}^{post-insp}},
\label{imrc1}
\end{equation}
where subscript, `f' denotes merger remnant parameters, and `insp' and `post-insp' correspond to the estimates coming from the low and high-frequency regimes, respectively. $\bar{M_{f}}$ and $\bar{\chi_{f}}$ are the symmetric combinations of $M_{f}$ and $\chi_{f}$ estimates from inspiral and post-inspiral regimes.
A similar expression can be written for dimension-less spin parameter,
\begin{equation}
\delta \chi_{f}=\frac{\Delta \chi_{f}}{\bar{\chi_{f}}}=2\frac{\chi_{f}^{insp}-\chi_{f}^{post-insp}}{\chi_{f}^{insp}+\chi_{f}^{post-insp}}.
\label{imrc2}
\end{equation}

{In} principle, these fractional deviations vanish if the data is consistent with GR{\endnote{Neglecting the instrumental noise, statistical fluctuations, and waveform model uncertainties, etc. These effects can lead to an offset from zero.}}. However,  one must perform a detailed parameter estimation analysis and estimate the statistical confidence that the GR deviation vanishes. This is illustrated in Figure~\ref{fig:imrct1} and see Reference~\cite{Ghosh:2016qgn} for more details of the method and demonstration on simulated binary signals. See References~\cite{Carson:2019kkh, Carson:2019yxq, Carson:2019rda} for studies projecting the possibilities of IMR consistency tests from combining information from current and future detectors. Especially, Reference~\cite{Carson:2019yxq} found that the multi-band observations can improve the constraints by a factor of 1.7.

\begin{figure}[H]
\includegraphics[scale=0.4]{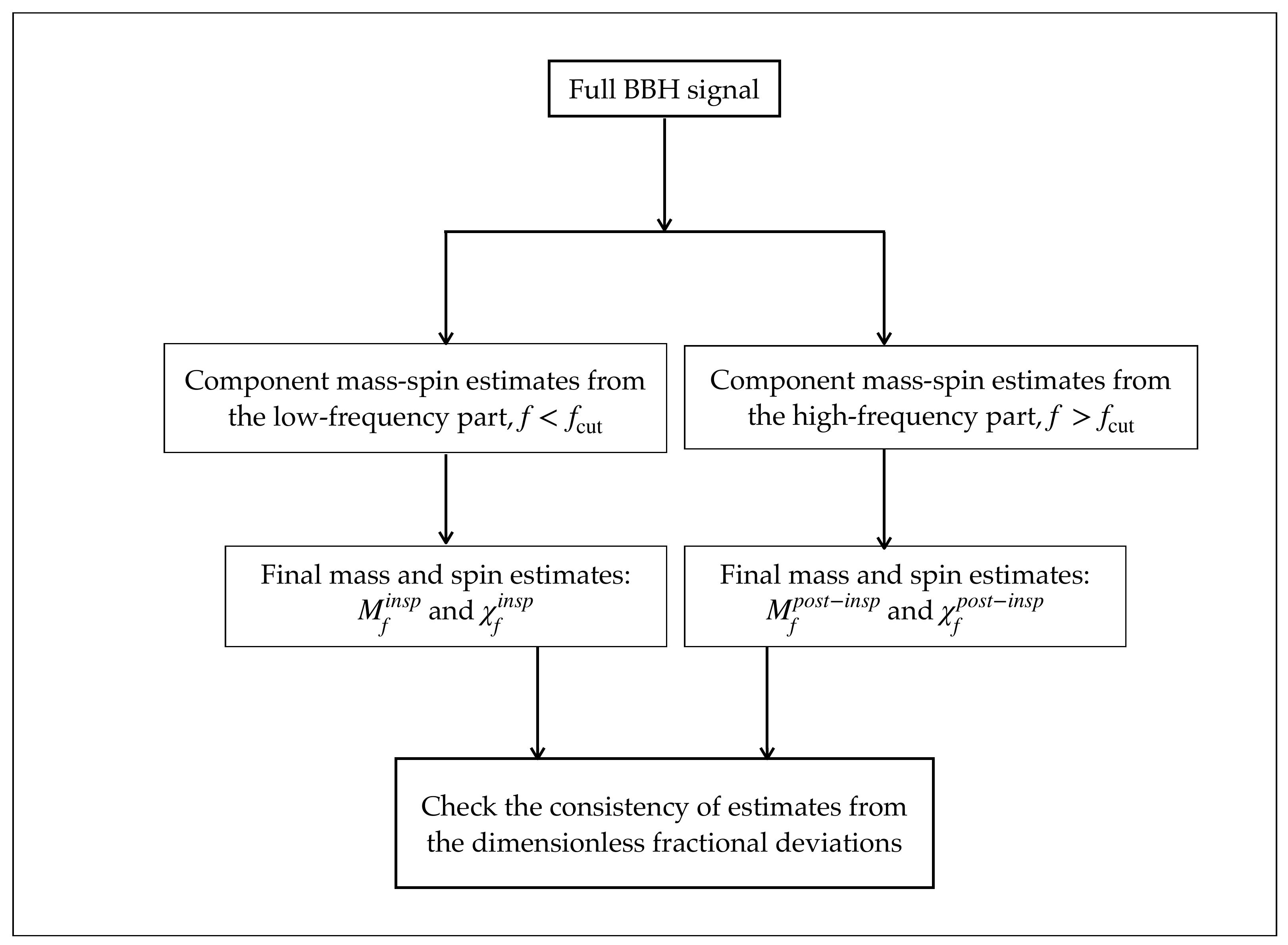}
\caption{{The} figure shows the graphical demonstration of the inspiral-merger-ringdown consistency test. We start with the full BBH signal. Analyze the signal by restricting the noise-weighted integral in the likelihood calculation to frequencies below and above the cut-off frequency, $f_{\rm{cut}}$. From the two independent mass-spin estimates obtained above, the merger remnant properties are calculated by averaging NR-calibrated final state fits. If the data is consistent with GR, the deviation parameters defined in Equations~(\ref{imrc1}) and~(\ref{imrc2}) will be zero, assuming that the waveform model employed accurately models a BBH evolution in GR.  See Reference~\cite{Ghosh:2016qgn} for more details and examples. }
\label{fig:imrct1}
\end{figure}

In Reference~\cite{GWTC2-TGR}, from the list of events satisfying the detection criteria, based on the false alarm probability of each event and the requirement of enough SNR in both inspiral and post-inspiral regimes,   posteriors distributions on $\delta M_{f}$ and $\delta \chi_{f}$ are obtained assuming uniform priors on these parameters.  In terms of the two-dimensional GR quantile, $Q_{GR}$--is the fraction of the posteriors enclosed by the iso-probability contours that contain the GR value. Reference~\cite{GWTC2-TGR} reports {\tt GW190814} as the most consistent event with the  quantile $Q_{GR} = 99.9 \%$.  

From the hierarchical combining method, the hyperparameters describing the Gaussian distribution are estimated to be  $(\mu, \sigma)=(0.02^{+0.11}_{-0.09}, <0.17)$ for $\delta M_{f}$ and \linebreak\mbox{$(\mu, \sigma)=(-0.06^{+0.15}_{-0.16}, <0.34)$} for $\delta \chi_{f}$ within the 90\% confidence interval in Reference~\cite{GWTC2-TGR}. The details can be found in Figure~\ref{fig:imrct2}. Assuming that the fractional deviations take the same value for all events, a less-conservative combined 90\% confidence interval of $\frac{\Delta M_{f}}{\bar{M_{f}}}=0.04^{+0.08}_{-0.06}$  and $\frac{\Delta \chi_{f}}{\bar{\chi_{f}}}=-0.09^{+0.11}_{-0.08}$ obtained in Reference~\cite{GWTC2-TGR}. This analysis employed {\tt{IMRPhenomPv2}} or {\tt{IMRPhenomPv3HM}} waveform models depending upon the information about the higher-mode content present in the binary signal~\cite{Breschi:2019wki} and so far the analysis finds all events to be consistent with GR.

\begin{figure}[H]
\includegraphics[scale=0.55]{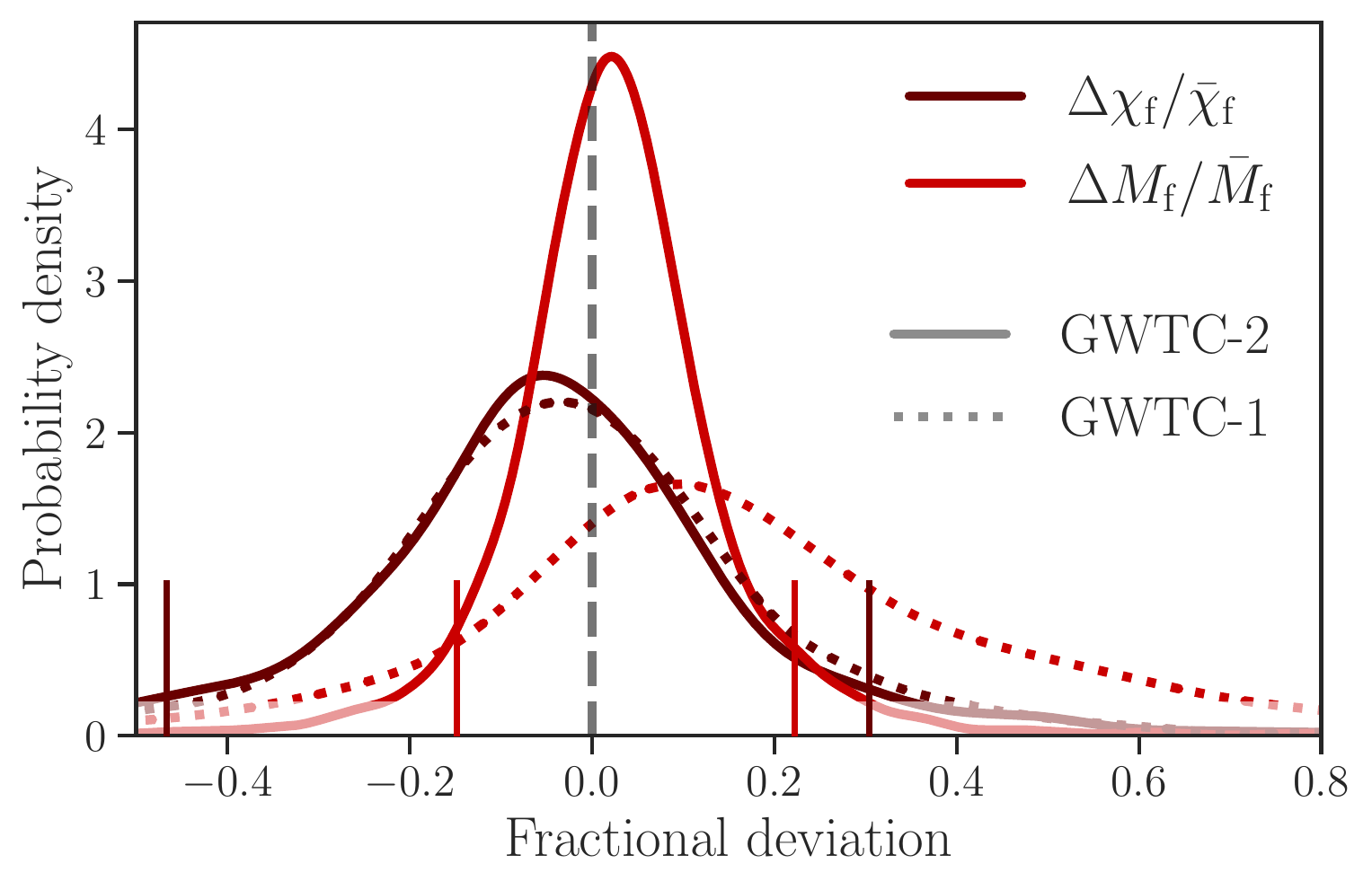}
\caption{{Posterior} distributions on the mass-spin deviation parameters (black and purple curves) were obtained by hierarchically combining the events~\cite{GWTC2-TGR}. Solid lines represent GWTC-2 events, while dotted lines represent GWTC-1 events. The vertical bars represent 90\% confidence intervals. }
\label{fig:imrct2}
\end{figure}

\subsection{Parametrized Tests of GR Based on Generation and Propagation of GWs}
\label{param_tests}

The parametrized tests are designed to capture any deviations from GR in the generation and propagation of GWs. This is achieved by introducing model-independent parametric variations in the gravitational waveform models and constrain those from the observed data. If GR is the correct theory, parametric deviations vanish, and the statistical bounds can be used to put constraints on the alternative theory models.

\subsubsection{Constraining the Parametrized Deviations from General Relativistic Inspiral-Merger-Ringdown Coefficients}

Any generic deviation from GR may modify the binary dynamics and its time evolution. This leads to measurable modifications to the equation of motion through the energy and angular momentum of the source and the flux. However, the inspiral-merger-ringdown dynamics are uniquely determined and well studied in GR through various techniques such as post-Newtonian theory, numerical relativity, and BH perturbation theory once we fix the intrinsic parameters of the binary system~\cite{Husa:2015iqa,PhysRevLett.111.241104,Hinder_2013,PhysRevLett.99.181101,PhysRevLett.96.111102,PhysRevLett.95.121101,PhysRevLett.96.111101,Blanchet:2008je,DAMOUR2001147,PhysRevLett.93.091101,PhysRevLett.74.3515}.

The inspiral coefficients are modeled analytically using post-Newtonian (PN) theory, which finds perturbative solutions to the binary evolution in terms of a velocity parameter, $v/c$, in the slow-motion limit ($v<<c$, $v$ is the PN parameter and $c$ is the velocity of light). Parametrized tests based on inspiral coefficients are investigated in detail~\cite{AIQS06a,Yagi:2011xp,AIQS06b,Yunes:2009ke,MAIS10,TIGER2014, PhysRevLett.74.3515,Blanchet:2005tk,Kidder:1995zr,DAMOUR2001147,Blanchet:2006gy,Arun:2008kb,Bohe:2012mr,Marsat:2012fn,Bohe:2013cla,Blanchet:2013haa,Damour:2014jta,Bohe:2015ana} and also demonstrated the applicability of the test using Bayesian framework ~\cite{TOGGW150914,O1BBH,GW170104,LIGOScientific:2018dkp,Yunes:2016jcc} in the past. Moreover, the possibility of constraining these parameters employing multiband observations has been studied in References~\cite{Carson:2019kkh,Carson:2019yxq,Carson:2019rda,Gupta:2020lxa}.
For an inspiralling compact binary system, the GW waveform can be schematically represented in the frequency domain as,
\begin{equation}
\tilde{h}(f) = A(f)\,e^{i\phi(f)},
\end{equation}
where $A(f)$ denotes the amplitude and ${\phi(f)}$ is the phase of the signal. Now, we introduce parametric deviations of the form, $\phi_{i}(f)\rightarrow (1+ \varphi_{i})\phi_{i}$. If GR is the correct theory, $ \hat{\phi}_{i}$ vanishes for the $N${th} PN order, where $i=N/2$ denotes the $N${th} PN order. The parametrized tests for post-Newtonian coefficients (pPN analysis) measure these deviation parameters and are one of the main tests of GR analyses performed for the detected binary signals so~far.

For a generic binary system, one needs to put bounds on the inspiral-merger-ringdown parametrized deviation coefficients separately. In this case, a relative deviation is introduced to each coefficient appearing through the  inspiral-merger-ringdown regimes as,
\begin{equation}
p_{i}\rightarrow(1+\delta{p_{i}})p_{i}.
\end{equation}

{The} set of free parameters, $\delta{p_{i}}$ include the inspiral
coefficients $\{ {\varphi}_{i}\}$ and post-inspiral coefficients
$\{{\alpha}_i,{\beta}_i\}$~\cite{Meidam:2017dgf,TIGER2014,LiEtal2011}. It is not
plausible to represent the post-inspiral coefficients analytically and they are
obtained by numerical fits.

The {\tt{IMRPhenomPv2}} waveform model~\cite{Husa:2015iqa,Khan:2015jqa,Schmidt:2010it} describes a precessing binary system in frequency domain with inspiral coeffiecients determined by PN theory and  intermediate, and merger–ringdown regions by finding appropriate  numerical fits. The transition frequency between the inspiral to merger-ringdown is defined as $GM(1+z)f_{c}^{PAR}/c^3 = 0.018$ ($M$ is the total mass of the binary system and $z$ is the redshift to the binary). 
The results for parametrized tests for post-Newtonian coefficients reported in Reference~\cite{GWTC2-TGR} relied on  the {\tt{IMRPhenomPv2}} waveform model which allows for parametrized deviations of the phenomenological coefficients describing the inspiral, intermediate ${\beta}_i = \{{\beta}_1, {\beta}_2\}$,  and merger-ringdown ${\alpha}_i = \{{\alpha}_2, {\alpha}_3, {\alpha}_4\} $ regions.

There is also another equally accepted method to perform the same analysis, which is not based on any particular waveform model but rather on theory-agnostic modifications applied to the inspiral coefficients of any waveform model. These low-frequency modifications (inspiral-only modifications $\{ {\varphi_{i}}\}$) are tapered to zero at high frequencies. That is, as the frequency reaches post-inspiral regions these modifications vanish and the signal agrees to a BBH signal described in GR.  This test is carried out employing {\tt SEOBNRv4\_{ROM}} waveform model in References~\cite{LIGOScientific:2018dkp,Bohe:2016gbl,Purrer:2015tud}. One of the main differences between the two approaches is that the deviations are applied to only the non-spinning coefficients for the first method. Still, for the second method, the aligned-spin waveform coefficients are modified. Both approaches provide consistent results when we compare them.

Denoting $\{ \varphi_{i}\}$ as the deviation from $N=i/2$ PN order, the list of inspiral coefficients we can put bounds from the data are,
\begin{equation}
\{  \varphi_{-2},   \varphi_{0},   \varphi_{1},   \varphi_{2},   \varphi_{3},   \varphi_{4},   \delta\varphi_{5},   \varphi_{6},   \varphi_{7}\}.
\end{equation}

{This} means the coefficients up to 3.5PN order ($i=7$) are available, including the logarithmic terms at 2.5 and 3PN orders. Due to its degeneracy with the coalescence phase, one cannot constrain the coefficient at 2.5 PN (term having no logarithmic dependence). Notice also that ${\varphi_{-2}}$ is zero in GR and, so as ${\varphi_{1}}$, it is introduced to account for specific alternative theories of gravity, especially those that predict dipolar radiation. For the two coefficients, ${\varphi_{1}}$ and ${\varphi_{-2}}$, we have the absolute deviations while all other parameters provide relative deviations from the respective GR coefficients. The absolute deviations are the differences between the true/actual value and the measured value, whereas the relative deviations are the ratio of absolute deviations to the true/actual value.

Among all the coefficients listed above, the best combined bound is obtained for the Newtonian coefficient in Reference~\cite{GWTC2-TGR}, $| {\varphi_{0}}|\leq4.4\times10^{-2}$ (neglecting the {$-1$}PN coefficient, $\varphi_{-2}$). Notice that this bound is weaker than the bounds from the double pulsar measurement by a factor of $\sim$3. Reference~\cite{GWTC2-TGR} reports that the posterior on $\delta\hat{p_{i}}$ is consistent with the GR prediction within the 90\% confidence interval for all the events considered. From the hierarchical analysis, the tightest (loosest) bound obtained is \linebreak$\varphi_{-2}=0.97^{+4.62}_{-4.07}\times10^{-3}$ ($\varphi_{6\ell}=0.42^{+1.67}_{-1.50}$). For both the circumstances, the GR hypothesis is preferred with quantiles $Q_{GR}=68\%$ ($Q_{GR}=69\%$, which is close to the median values for both the cases. Figure ~\ref{fig:param} shows the combined posterior distributions obtained from the GWTC-2 events considering both the hierarchical method and a restricted method, where the deviation parameters do not allow for variance between events. Results from both the studies described above are found to be consistent with GR.

\end{paracol}
\nointerlineskip

\begin{figure}[H]
\widefigure
\includegraphics[scale=0.55]{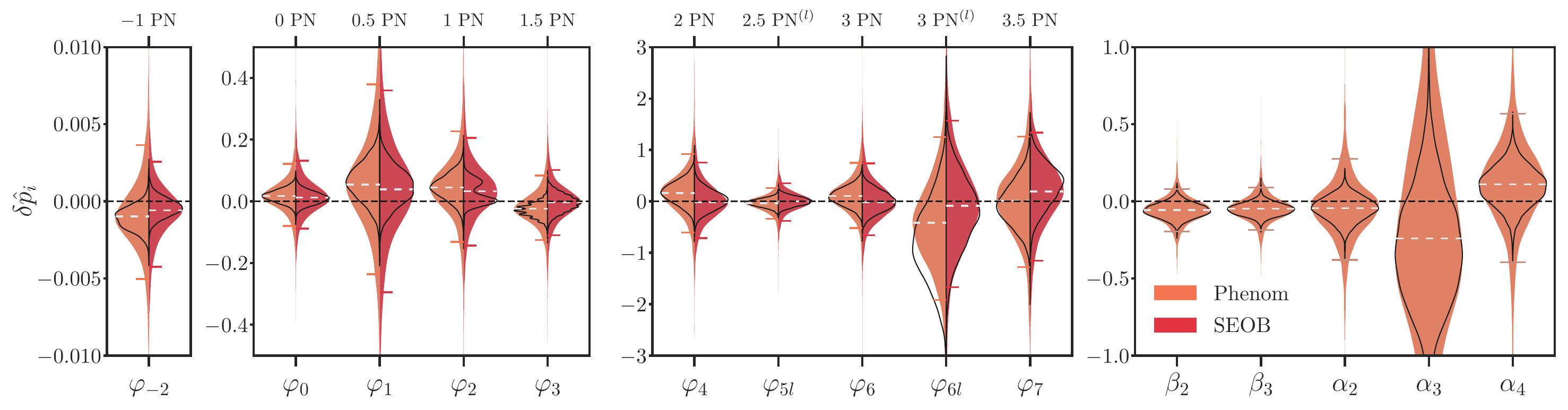}
\caption{{Posterior} distributions on the inspiral $\{\varphi_{i}\}$ and the post-inspiral coefficients $\{{\alpha}_i,{\beta}_i\}$ considering the detected BBH population through first, second and (first half of) third observing runs~\cite{GWTC2-TGR}. The filled posteriors are obtained from hierarchical analysis and the unfilled black distributions assuming that all events share a common value for the deviation parameter instead of independently varying it for each event. The results in orange are obtained from the phenomenological waveform model {\tt{IMRPhenomPv2}}, while in red are from {\tt SEOBNRv4\_{ROM}} waveform model. Error bars represent 90\% confidence intervals for the hierarchical results, and the white dashed line denotes the median. The expected GR value, $ \delta{p_{i}}=0$ is shown by dashed horizontal line.}
\label{fig:param}
\end{figure}
\begin{figure}[H]
\widefigure
\includegraphics[scale=0.45]{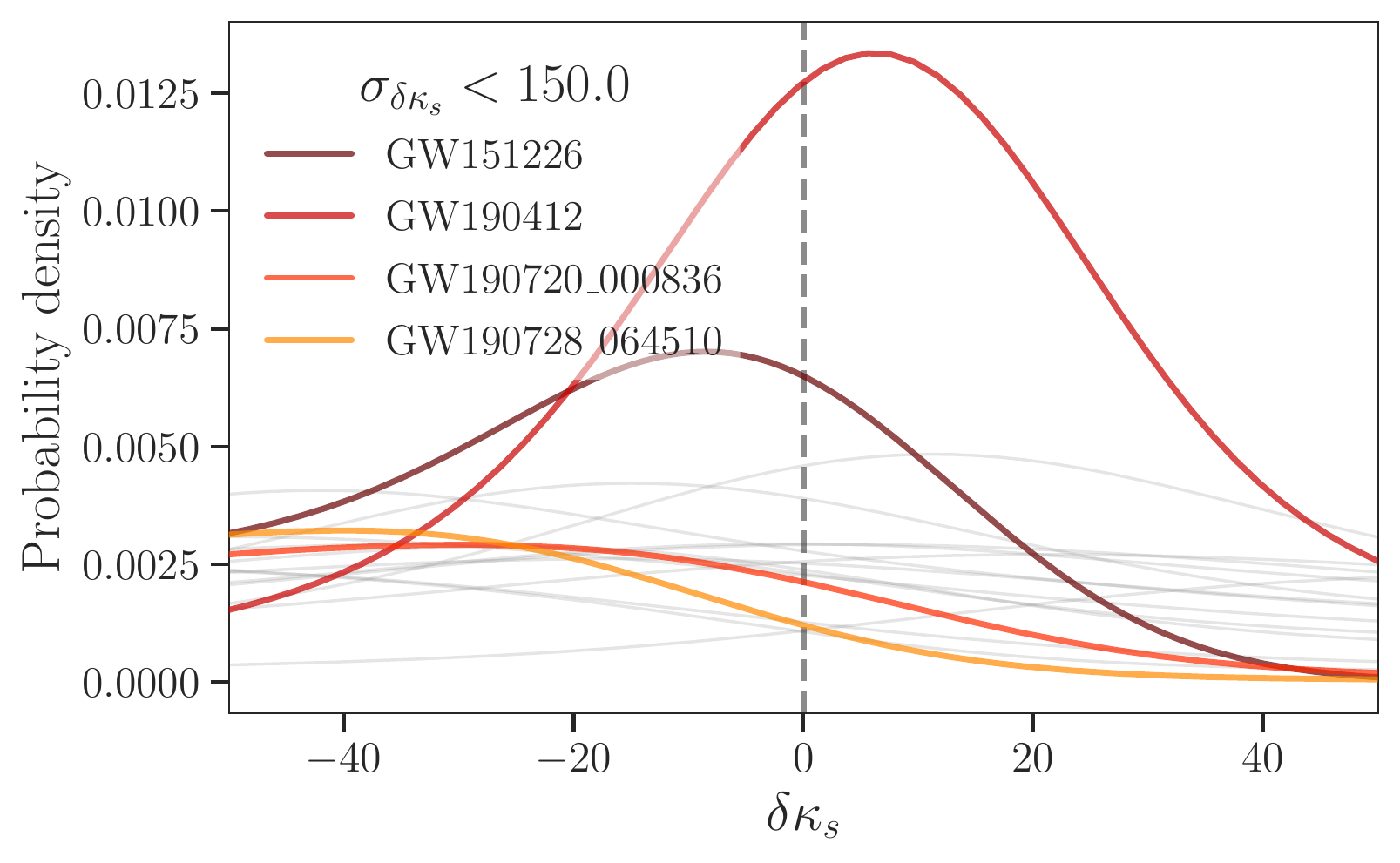}
\includegraphics[scale=0.45]{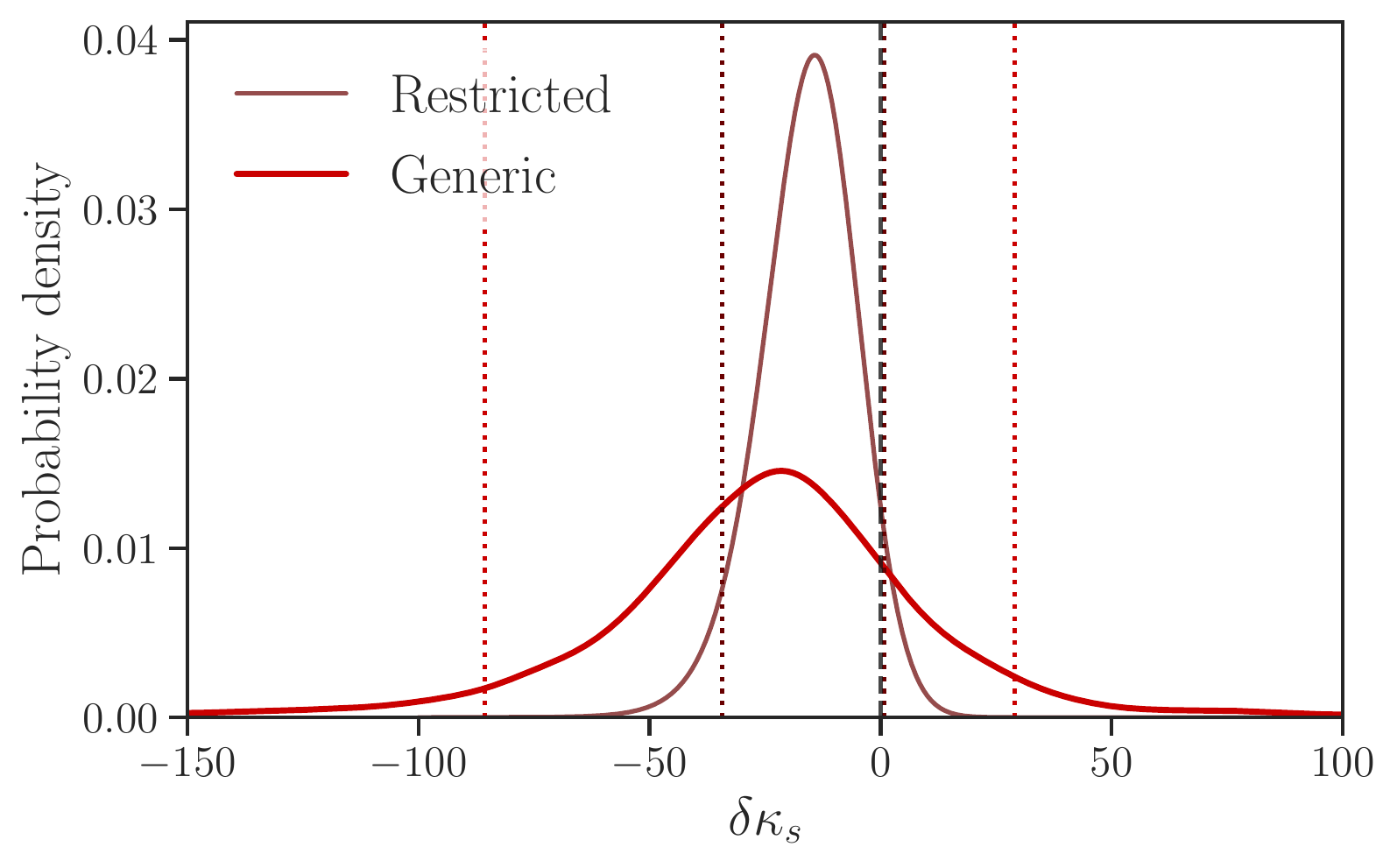}
\caption{\textbf{Left}: {Posterior} distributions on the inspiral $\delta\kappa_s$ for selected events detected through the first, second, and third observing runs of LIGO-Virgo detectors~\cite{GWTC2-TGR}. The events highlighted give better constraints on the $\delta\kappa_s$ parameter and all the other events considered are shown in grey. The vertical line at $\delta\kappa_s=0$ indicates the BBH value. \textbf{Right}: The bounds obtained on the $\delta\kappa_s$ parameter from combining information from multiple events. The restricted method assumed universal values of $\delta\kappa_s$ for all the events. The generic way allows the possibility of varying $\delta\kappa_s$ among different events according to a Gaussian distribution whose characteristics are obtained from the data.  }
\label{fig:sim}
\end{figure}

\unskip
\begin{paracol}{2}
\switchcolumn

\subsubsection{Tests of BBH Nature from Spin-Induced Quadrupole Moment Measurements}
\label{sim}
Spin-induced multipole moments arise due to the spinning motion of the compact object and take unique values for BHs given mass and spin. For the BBH signals, these effects are included in the post-Newtonian modeling, and they appear along with the spin-spin terms and can be schematically represented as,
\begin{equation}
Q = -\kappa\, \chi^{2}\,m^3,
\label{eq:sim}
\end{equation}
here $\kappa$ is the spin-induced quadrupole moment coefficient, $m$ is the mass, and $\chi$ is the dimensionless spin parameter. For BHs, $\kappa_{BH}=1$ from the no-hair conjecture \cite{Hansen74, Carter71, Gurlebeck:2015xpa} and for any other compact objects, the value may vary depending upon the properties of the star. Through numerical relativity simulations of slowly spinning neutron stars, it is found that the value of $\kappa$ varies between $\kappa_{NS}=$ 2 and 14  \cite{Pappas:2012qg, Pappas:2012ns, Harry:2018hke}. On the other hand, for more exotic stars like boson stars, the value of $\kappa$ can be even larger and found to vary between $\sim$10 and 100 \cite{Ryan97b, Herdeiro:2014goa}. 

It has been shown that one can introduce parametrized deviations of the form, \mbox{$\kappa=(1+\delta\kappa)$} and put bound on  $\delta\kappa$~\cite{Krishnendu:2019tjp,Brown:2006pj,Rodriguez:2011aa,Barack:2006pq,Babak2017}. An inspiralling binary system is parametrized by two such parameters, corresponding to both the binary components, $\delta\kappa_1$ and $\delta\kappa_2$. Simultaneous measurement of both these parameters  will end up giving weak constraints on either parameter, hence it is proposed to measure $\delta\kappa_s=0.5(\delta\kappa_1+\delta\kappa_2)$ keeping $\delta\kappa_a=0.5(\delta\kappa_1-\delta\kappa_2)=0$. This is a safe assumption if we are testing the  BBH nature of the detected signal~\cite{Krishnendu:2017shb,Krishnendu:2018nqa}. The analysis performed on the first, second, and third observing runs of LIGO-Virgo detectors provided good constraints on events with non-zero spins which include {\tt GW151226}, {\tt GW190412}, {\tt GW190720\_000836}, and {\tt GW190728\_064510}. Employing {\tt{IMRPhenomPv2}} waveform model~\cite{Husa:2015iqa,Khan:2015jqa,Schmidt:2010it} and assuming prior distribution on $\delta\kappa_s$ ranging uniform between [$-$500, 500]~\cite{GWTC2-TGR}.  These events, where the posteriors are very different from the prior knowledge,  are highlighted in Figure~\ref{fig:sim}.


\end{paracol}
\nointerlineskip
\pagebreak

\begin{paracol}{2}
\switchcolumn

With the restricted assumption that $\delta\kappa_s$ take the same value for all events, Reference~\cite{GWTC2-TGR} reported a combined bound on $\delta\kappa_s$ within the $90\%$ confidence interval as $\delta\kappa_s=-15.2^{+16.9}_{-19.0}$. Using the hierarchical analysis, the hyperparameters are constrained to $\mu=-24.6^{+30.7}_{-35.3}$ and $\sigma<52.7$ with $\delta\kappa_s=-23.2^{+52.2}_{-62.4}$ and which is again consistent with the null \mbox{($\mu=\sigma=0$)} hypothesis at $90\%$ confidence interval. The hypothesis stating that the population contains all BBHs is favored by the population containing all the non-BBH hypothesis by a combined Bayes factor of 11.7. The analysis found that the data are consistent with the BBH hypothesis.

\subsubsection{Tests of Gravity from GW Propagation}
\label{prop}

GW propagation in GR is non-dispersive and described by the dispersion relation,
\begin{equation}
E^2=p^2c^2.
\label{gr_disp}
\end{equation}

{Equivalently}, the velocity of propagation of GWs in GR is independent of the frequency of the radiation. As a consequence, the graviton is massless with a corresponding infinite Compton wavelength. There are alternative theories of gravity predicting GWs with dispersion where the local Lorentz invariance is not respected~\cite{Mattingly2005ModerntestsofLorentzinvariance}.

For a generic theory of gravity, the GR dispersion relation may require modifications and the following equation can account for such propagation effects~\cite{AnuradhaArun2017,AmelinoCamelia2002DoubleSpecialRelativity,Will2014LivingReviewsinRelativity,Calcagni2009FractionalUniverseandQG},
\begin{equation}
E^2=p^{2}\,c^2+\mathcal{A}\,p^{\alpha}\,c^{\alpha}.
\label{mod_disp}
\end{equation}

{We} can re-parametrize the gravitational waveforms so that they also account for the propagation effects given in Equation~(\ref{mod_disp}).  In Equation~(\ref{mod_disp}),  $\mathcal{A}$ is the dispersion amplitude and has dimension of $[Energy]^{2-\alpha}$, and $\alpha$ is a dimensionless constant.  These paramtetrized modifications can be constrained from the data and these bounds can be translated into constraints on different alternative gravity models. For example, $\alpha=0$ and $\mathcal{A}>0$  correspond to massive graviton theories~\cite{Will2014LivingReviewsinRelativity}, $\alpha=2.5$  corresponds to multifractional spacetime \cite{Calcagni2009FractionalUniverseandQG}, and $\alpha=3$  corresponds to double special relativity~\cite{AmelinoCamelia2002DoubleSpecialRelativity}, etc.

As shown in References~\cite{AnuradhaArun2017, Mirshekari:2011yq}, one can use GW observations to get constraints on the modified dispersion parameters. The first bound on the Compton wavelength (which has a finite value for any massive graviton theory) from GW observations of a BBH signal is $\lambda_{g}>10^{13} ~\rm{km}$~\cite{TOGGW150914} and this has been extended to more generic cases in the subsequent analyses~\cite{GW170104, TGRGWTC-1, GW170814}. This bound on the Compton wavelength translates to a graviton mass, $m_{g}\leq5\times 10^{-23}~\rm{eV/c^{2}}$, and this is a stronger bound compared to the solar system constraints~\cite{TGRGWTC-1}. From the GWTC-2 data~\cite{GWTC2-TGR}, a factor of 2.7 improvements is observed on this bound, and the graviton mass bound correspondingly changes to $m_{g}\leq1.76\times10^{-23}~\text{eV/c}^2$ with $90\%$ credibility.


Note that the past studies have investigated the possibility of dispersion of GWs described by GR due to specific physical effects.  For example, the nonlinear interaction between charged particles and GWs may lead to dispersion of GWs when GWs pass through astrophysical plasma in the presence of magnetic fields~\cite{Macedo:1983wcr,Gangopadhyay:2014dpa,1992A&A...261..664V,Grishchuk:1981fp,1993A&A...275..309K,1995A&A...294..313K,1996,JOUR,Wickramasinghe:2015zja,Revalski:2015aka}.

\subsection{Tests Based on the Merger Remnant Properties}
\label{rem_properties}

The merger remnant of a BBH system emits GWs to settle down to the stationary state, and this distinct signal from stellar-mass BBHs can be measured using the current GW detectors. Consequently, many tests for remnant nature have been proposed and performed on the detected GW events. We briefly discuss GW tests based on merger remnant properties here.

\subsubsection{No-Hair Theorem Based Tests from the Quasi-Normal Mode Ringdown Radiation (BH~Spectroscopy)}

According to GR, the remnant formed after a BBH coalescence is a perturbed Kerr BH, and this BH attains the stationary state by emitting GWs. This damped sinusoidal signal (BH ringdown radiation) is characterized by quasi-normal-modes (QNMs) with frequency $f$ and damping time $\tau$. Both the damping time and frequency of this oscillation are determined by the mass and spin of the Kerr BH formed after the merger ($M_f$ and \mbox{$\chi_f$)~\cite{Kokkotas1999QNMs, Carter71, PhysRev.164.1776, Hawking,PhysRevLett.34.905, Mazur_1982}}. In GR, the ringdown waveform is a superposition of damped sinusoids and takes the form,
\end{paracol}
\nointerlineskip
\vspace{6pt}

\begin{equation}
\label{eq:ring_wf}
\begin{aligned}
h_{+}(t) - i h_{\times}(t) = \sum_{\ell = 2}^{+\infty} \sum_{m = - \ell}^{\ell} \sum_{n = 0}^{+\infty} \; & \; \mathcal{A}_{\ell m n} \; \exp \left[ -\frac{t-t_0}{(1+z)\tau_{\ell m n}} \right] \exp \left[ \frac{2\pi i f_{\ell m n}(t-t_0)}{1+z} \right] {}_{-2}S_{\ell m n}(\theta, \phi, \chi_{\rm f}), \\
\end{aligned}
\end{equation}
\begin{paracol}{2}
\switchcolumn
\noindent where $z$ is the cosmological redshift, and the $(l, m, n)$ indices label the QNMs ($(\ell, m)$ are the angular multipoles, whereas $n$ is the order of modes given $(\ell, m)$. All the $f_{\ell m n}$ and $\tau_{\ell m n}$ are determined by the final mass and spin of the binary system (this is called the final state conjecture). For a perturbed Kerr BH, the damping time and frequency of each quasi-normal-mode can be calculated from BH perturbation theory as a function of its final mass and spin~\cite{Chandra70PRL,1973ApJ...185..635T,1971ApJ...170L.105P,PhysRevD.1.2870}. Assuming that the mergers we observe are BBHs, from the independent $M_f$ and $\chi_f$  post-merger measurements we can test the final state conjecture (commonly known as   the BH spectroscopy)~\cite{PhysRevD.99.123029,Isi:2019aib,Dreyer:2003bv,Berti:2005ys,Gossan:2011ha,Healy:2016lce,Carullo:2018sfu,Brito:2018rfr,Bhagwat:2019bwv,Bhagwat:2019dtm,Cabero:2019zyt}. The complex amplitude $ \mathcal{A}_{\ell m n}$ is a measure of the mode excitation and the phase of these modes at a reference time~\cite{Kamaretsos:2012bs,London:2014cma,London:2018gaq}.

{\tt{PYRING}} is a toolkit to perform BH spectroscopy which is completely implemented in the time domain~\cite{PhysRevD.99.123029, Isi:2019aib}. As both templates and the likelihood are modeled in the time domain, spectral leakage is reduced~\cite{Cabero:2017avf}. Mainly assumed template models for this study are, $Kerr_{220}$ ($\ell=|m|=2, n=0$ contributions of Equation~(\ref{eq:ring_wf})), $Kerr_{221}$ ($\ell=|m|=2$, \mbox{$n=0, 1$} contributions of Equation~(\ref{eq:ring_wf})), and $Kerr_{HM}$ (all fundamental prograde modes with $\ell\leq4, n=0$ contributions of Equation~(\ref{eq:ring_wf}) and also taking into account mode-mixing~\cite{London:2018gaq}). The frequencies and damping times are predicted in terms of final mass and spin for all these cases. The remnant quantities, $M_f$ and $\chi_f$, are estimated assuming uniform priors on these parameters. We do not consider the higher overtones (n > 1) as those are not expected to provide constraints with the current sensitivity of detectors.

Another equally established technique for the QNM analysis is the parametrized-{\tt SEOBNRv4HM} ({\it{pSEOB}}) analysis, which employs a parametrized version of the EOB waveform model~\cite{Cotesta:2018fcv} accounting for aligned spins and higher modes~\cite{GWTC2-TGR, Ghosh:2021mrv, Brito:2018rfr}. The {\it{pSEOB}} analysis differs from {\tt{PYRING}} in that it measures the ringdown frequency within a complete IMR waveform model framework using the full SNR of the signal. It is not dependent on a ringdown time definition. In this framework, one parameterizes the frequency and damping time of the $\ell=m=2$ by introducing a fractional deviation from the nominal GR prediction and constrains these fractional deviation parameters directly from the data. That is, $f_{220}=f_{220}^{GR}(1+\delta\hat{f}_{220})$ and $\tau_{220}=\tau_{220}^{GR}(1+\delta\hat{\tau}_{220})$~\cite{Cotesta:2018fcv}, where $f_{220}^{GR}$ and $\tau_{220}^{GR}$ are frequency and damping time if GR is the correct theory of gravity.
From both these approaches, by performing detailed analyses on the GWTC-2 events, no indication of the presence of non-BH behavior was reported~\cite{GWTC2-TGR}.

\subsubsection{Testing the Nature of Merger Remnant from the Measurement of Late Ringdown~Echoes}

One can ask the question as to if the merger remnant is not a BH but instead an exotic compact object (ECO) with a light-ring and reflective surface, instead of an event horizon as in the case of BHs~\cite{Abedi:2016hgu, Tominaga:1999iy, Lunin:2002qf, Lunin:2001jy}. For these hypothetical cases, the GWs can be trapped in between the effective potential at the centre and the reflective surface, leading to the emission of GWs as a train of repeating pulses known as GW echoes. BHs produce no echo signal as there is no possibility of an out-going boundary condition at the BH event horizon. GW echoes are unique probes of any non-BH compact object formation (especially exotic objects like gravastars, fuss balls) after the binary coalescence~\cite{Maggio:2020jml,Mazur:2001fv, McManus:2019ulj,Abedi:2016hgu}. For an illustration of GW echoes originating from a binary merger, we point out \mbox{Figure~\ref{fig:echo}} (also see Figure 2 of Reference~\cite{Abedi:2016hgu} for the original {\tt GW150914} signal on top of the best-fit echo template from a template-based echo analysis).

\begin{figure}[H]
\includegraphics[scale=0.3]{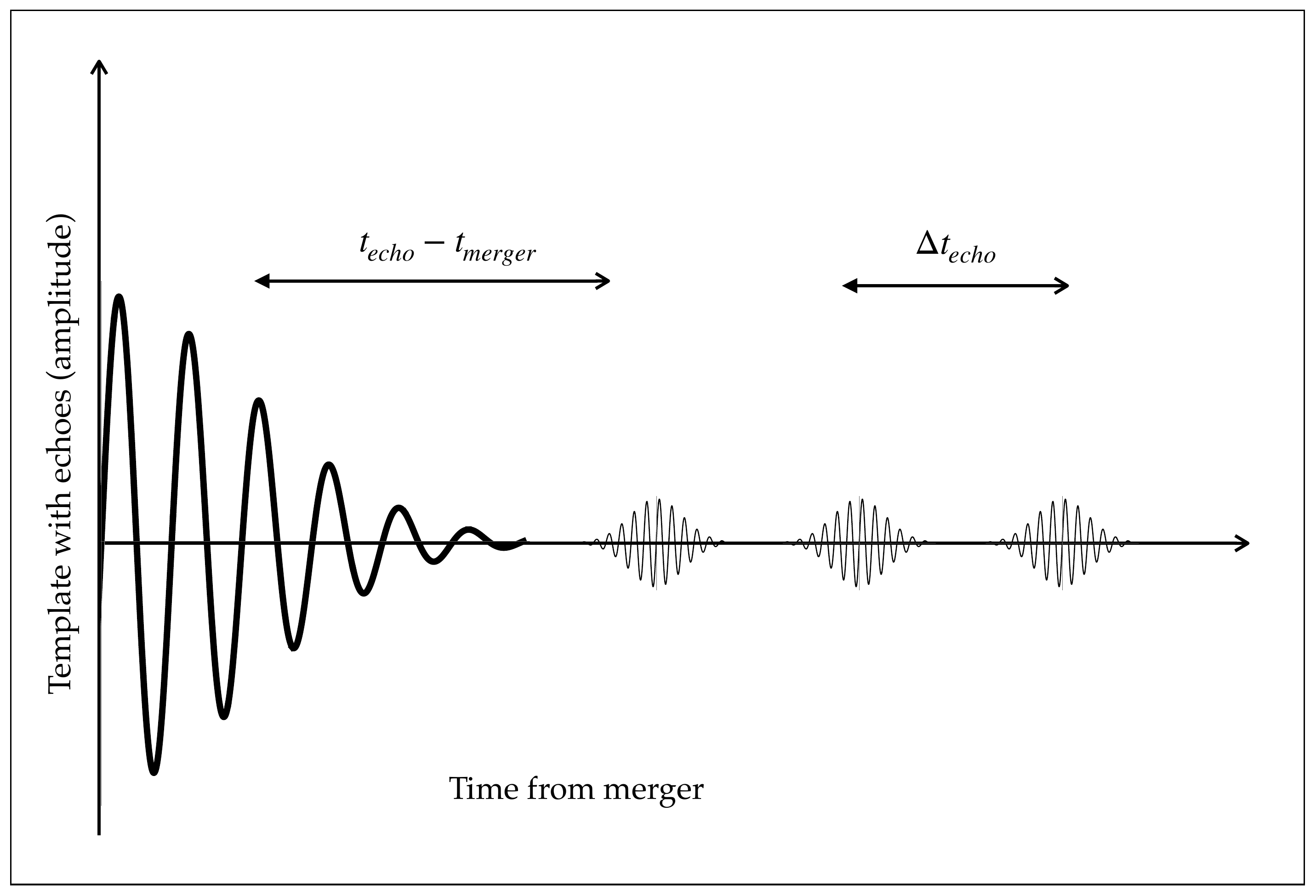}
\caption{{Figure} demonstrating an echo template. Here $\Delta t_{echo}$ denotes the time-delay between two consecutive echoes; also, the horizontal line shows the time delay between binary merger and the first echo.  (In Reference~\cite{Abedi:2016hgu}, the time domain template along with the best-fit echoes template for {\tt GW150914} from a template-based echo analysis is plotted.) }
\label{fig:echo}
\end{figure}

In the template-based framework, the echo signal is modeled with five extra parameters, characterizing the echo: the relative amplitude of the echoes, the damping factor between each echo, the start time of ringdown, the time of the first echo concerning the merger, and the time delay between each echo ($\Delta t_{echo}$ in Figure~\ref{fig:echo}). Reference~\cite{Abedi:2016hgu} studies this method and subsequent discussions on {\tt GW150914} in detail. For the GWTC-2 events~\cite{GWTC2-TGR}, assuming uniform priors on each of these echo parameters and employing the {\tt{IMRPhenomPv2}} waveform model (except for the case {\tt GW190521}, where {\tt Sur7dq4}, is a surrogate model for precessing BBH system directly interpolates the numerical relativity waveforms, is used), a Bayesian analysis is performed to investigate the evidence for echoes. Bayes factor $B_{IMR}^{IMRE}$ (comparing the two hypotheses $IMRE$ data best fits an echo model, and data best fits a model without echoes $(IMR)$) are computed for each event. The data did not show evidence for echoes~\cite{GWTC2-TGR}, except for the event {\tt GW190915\_235702}, which showed the highest value for $B_{IMR}^{IMRE} = 0.17$ indicating a negligible evidence for echoes~\cite{GWTC2-TGR}. Reference~\cite{GWTC2-TGR} reports that the posteriors on the echoes parameters returned their prior distribution, pointing to a null detection of GW echoes. 

\subsection{Constraints on the Polarization States of GWs}
\label{pol}

Generic metric theories of gravity predict six independent degrees of freedom for the metric tensor, which can be identified as polarization states of GWs. More than the two tensor (spin 2) degrees of freedom allowed by GR, there is a possibility of two vectors (spin 1) degrees of freedom and two scalars (spin 0) degrees of freedom~\cite{Will2014LivingReviewsinRelativity} in such cases. Out of these six modes, three of them are transverse, and three are longitudinal. The first constraints on the polarization states of the GWs from observations are detailed in~\cite{TOGGW150914}. Still, the results were uninformative as the data from the two LIGO instruments are not enough to constrain the extra polarization mode. This is possible only if there are data available from one another detector with another orientation. Hence the polarization test was first demonstrated with the three-detector event {\tt GW170814}, and the improved results compared to Reference~\cite{TOGGW150914} are available in Reference~\cite{GW170814}. In Reference~\cite{GW170814} also the tensor modes hypothesis was favored over scalar and vector modes as shown in Reference~\cite{TOGGW150914}.

A detailed analysis was performed in Reference~\cite{GWTC2-TGR} considering all GW events observed till the first half of the third observing run of LIGO/Virgo detectors. Reference~\cite{GWTC2-TGR} reports the highest (lowest) Bayes factor for {\tt GW190720\_000836 }({\tt GW190503\_185404}) with \mbox{$\log_{10}B^{T}_{V}=0.139$}, $B^{T}_{S}=0.138)$ ($\log_{10}B^{T}_{V}=0.074$ and $B^{T}_{S}=-0.072)$, here $B^{T}_{V}$ and $B^{T}_{S}$ represent the Bayes factors for full tensor versus full-vector and full-scalar hypotheses respectively. The $B^{T}_{S}$ is slightly larger than $B^{T}_{V}$ and this is explained by the intrinsic geometry of the LIGO-Virgo antenna patterns~\cite{Nielsen:2018lkf}. One should also notice that any of these results do not account for the possibility of mixed polarization. This topic has to be explored in the future when more detectors become operational.

\section{Summary}
\label{con}

This review article provides a brief overview of the tests of GR performed during the first three observing runs of the LIGO-Virgo detectors, including tests of consistency with GR ({Section}~\ref{consistency_tests}), parameterized tests ({Section}~\ref{param_tests}), tests based on the merger remnant properties ({Section}~\ref{rem_properties}), and tests for GW polarizations ({Section}~\ref{pol}). Along with some technical details about each test, we also provide a short discussion pointing to the prospects of these various tests.

In this article, we only focused on signals which are consistent with BBHs. The detection of the first binary neutron star merger event opened up different possibilities of testing GR from combined electromagnetic and GW observations~\cite{GW170817, LIGOScientific:2018dkp}. Many tests we detailed here may have overlaps or redundant information which is not accounted for here. Though the instrumental noise mainly dominates the current measurement uncertainties, we cannot exclude the possibility of any systematic bias arising due to un-modeled effects present in the waveform models. Another critical point is that the tests discussed here are all model-independent tests (null tests). In other words, we are not assuming any alternative gravity theory models here, and every test is capturing the deviation from GR in a model-agnostic way.

\vspace{6pt}



\authorcontributions{{Conceptualization, investigation, visualization: N.V.K.; Writing: N.V.K. and F.O.; Review \& editing, supervision: F.O. All authors have read and agreed to the published version of the manuscript.}}

\funding{This research received no external funding.}



\dataavailability{{This research has made use of data, software and/or web tools obtained from the Gravitational Wave Open Science Center (\url{https://www.gw-openscience.org/} accessed on 8 December  2021), a service of the LIGO Laboratory, the LIGO Scientific Collaboration and the Virgo Collaboration. LIGO is funded by the U.S. National Science Foundation. Virgo is funded, through the European Gravitational Observatory (EGO), by the French Centre National de Recherche Scientifique (CNRS), the Italian Istituto Nazionale della Fisica Nucleare (INFN) and the Dutch Nikhef, with contributions by institutions from Belgium, Germany, Greece, Hungary, Ireland, Japan, Monaco, Poland, Portugal, Spain. The data presented in this study are openly available at \url{https://doi.org/10.7935/903s-gx73} accessed on 8 December  2021.}}

\acknowledgments{This work was supported by the Max Planck Society’s Independent Research Group Grant. We thank Ajit Mehta for carefully reading this article and providing comments. We are thankful to Angela Borchers Pascual and Anuradha Gupta for very useful comments and discussions. This document has LIGO preprint number {\tt LIGO-P2100349}. }

\conflictsofinterest{{The authors declare no conflict of interest.}}

\abbreviations{Abbreviations}{The following abbreviations are used in this manuscript:\\

\noindent
\begin{tabular}{@{}ll}
GW & Gravitational wave\\
BBH & Binary black hole\\
GR & General theory of relativity\\
NG & Newtonian gravity \\
GWTC & Gravitational-wave transient  catalog \\
O1/O2/O3 & First/Second/Third observing runs of LIGO/Virgo \\
LVK & LIGO-Virgo-KAGRA scientific collaboration \\
LVC & LIGO-Virgo scientific collaboration

\end{tabular}}

\appendixtitles{no} 
\begin{adjustwidth}{-5.0cm}{0cm}
\printendnotes[custom]
\end{adjustwidth}
\end{paracol}
\reftitle{References}

\end{document}